\documentclass[proof]{WileyASNA-v2}

\usepackage[utf8]{inputenc}
\usepackage{newunicodechar,graphicx}
\DeclareRobustCommand{\okina}{%
  \raisebox{\dimexpr\fontcharht\font`A-\height}{%
    \scalebox{0.8}{`}%
  }%
}
\newunicodechar{ʻ}{\okina}

\articletype{Article}

\received{01 March 2023}
\revised{10 July 2023}
\accepted{10 July 2023}

\begin{document}

\title{The Maunakea Spectroscopic Explorer: Thousands of Fibers, Infinite Possibilities}

\author[1]{Andrew Sheinis*}
\author[2]{Samuel C. Barden}
\author[3]{Jennifer Sobeck}
\author[4]{the MSE Team}

\authormark{Sheinis and the MSE Team \textsc{et al}}

\address[1]{\orgdiv{Maunakea Spectroscopic Explorer}, \orgname{Canada-France-Hawai\okina i Telescope)}, \orgaddress{\country{United States}}}

\corres{*Andrew Sheinis \email{sheinis@cfht.hawaii.edu}}

\presentaddress{65-1238 Mamalahoa Hwy, Kamuela HI 96743}

\abstract{
 The Maunakea Spectroscopic Explorer (MSE) is a massively multiplexed spectroscopic survey facility that will replace the Canada-France-Hawai\okina i Telescope over the next two decades. This 12.5-meter telescope, with its 1.5 square degree field-of-view, will observe 18,000 – 20,000 astronomical targets in every pointing from 0.36-1.80 $\mu$m at low/moderate resolution (R$\sim$3000, 6000) and from 0.36-0.90 $\mu$m at high resolution (R$ \sim$30,000). Parallel positioning of all fibers in the field and provide simultaneous full-field coverage for both resolution modes.  Unveiling the composition and dynamics of the faint Universe, MSE will impact nearly every ﬁeld of astrophysics across all spatial scales, from individual stars to the largest scale structures in the Universe, including (i) the ultimate Gaia follow-up facility for understanding the chemistry and dynamics of the distant Milky Way, including the distant halo at high spectral resolution, (ii) the unparalleled study of galaxy formation and evolution at cosmic noon, (iii) the determination of the neutrino mass, and (iv) the generation of insights into inﬂationary physics through a cosmological redshift survey that probes a large volume of the Universe with a high galaxy density. Initially, CFHT will build a Pathfinder instrument to fast-track the development of MSE technology while providing multi-object and IFU spectroscopic capability.
}


\maketitle

\section{Introduction}\label{sec1}
Highly multiplexed spectroscopic facilities have recently been or will soon be implemented on a variety of 4-meter class telescopes, e.g., WEAVE \citep{2012SPIE.8446E..0PD}, DESI \citep{2016arXiv161100036D}, and 4MOST \citep{2019Msngr.175....3D}.  These facilities will enable extensive scientific surveys exploring the nature of dark energy and the structure of the Milky Way Galaxy.  However, the depth of those surveys will be somewhat limited due to the aperture of their associated telescopes.  Current imaging surveys already produce fainter targets than these spectroscopic facilities can easily observe.  For example, the Large Survey of Space and Time (LSST) of the Vera C Rubin Observatory \citep{2019ApJ...873..111I} will soon flood the field with targets roughly six magnitudes fainter than the background sky, magnitudes that 4-m class facilities can not reach.  Hence the need for larger-aperture facilities to provide spectroscopic follow-up of a host of faint astronomical objects.

\begin{figure*}[th!]
\centerline{\includegraphics[width=140mm]{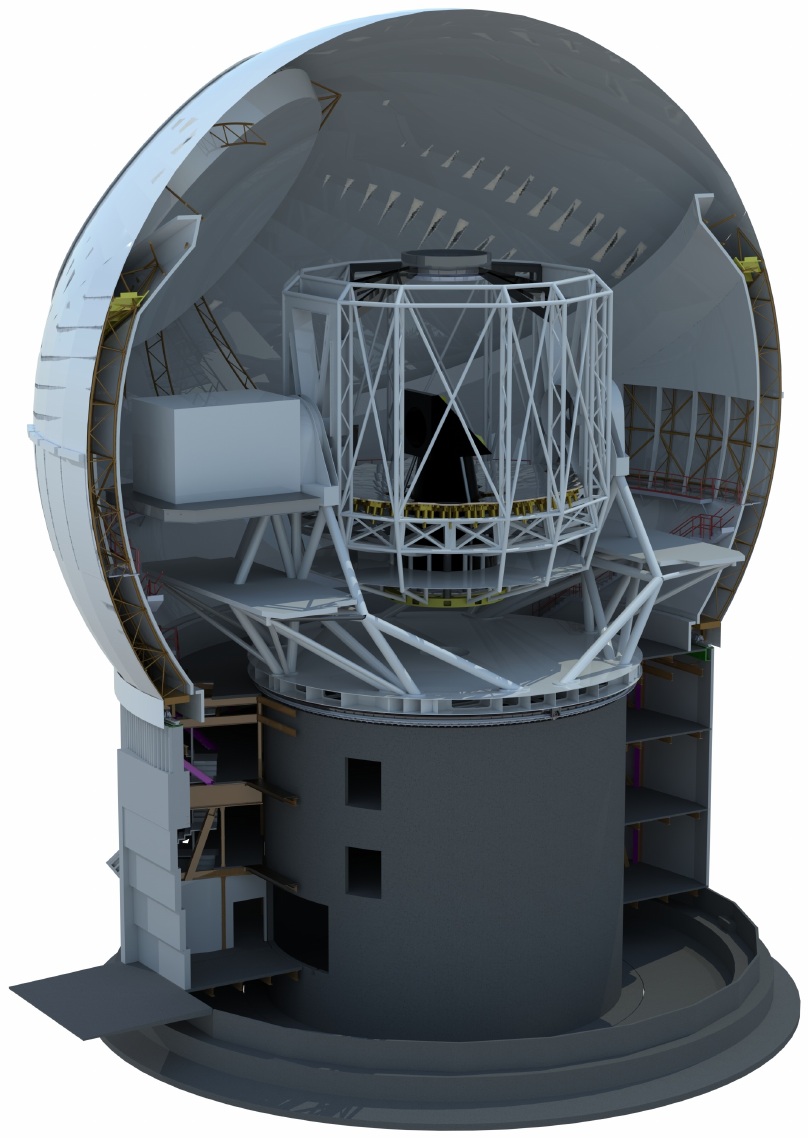}}
\caption{Rendering of the Maunakea Spectroscopic Explorer \label{fig1}}
\end{figure*}

The Maunakea Spectroscopic Explorer (MSE)\citep{2022SPIE12187E..1BB} will be a massively multiplexed spectroscopic survey facility and will replace the Canada-France-Hawai\okina i Telescope over the next two decades (Figure \ref{fig1}). This 12.5-meter telescope, with its 1.5 square degree field-of-view, will observe 18,000 – 20,000 astronomical targets in every pointing from 360-1800 nm at low/moderate resolution (R$\sim$ 3000, 6000) and from 360-900 nm at high resolution (R$\sim$30000). A  fiber positioner will enable parallel positioning of all fibers in the field and provide simultaneous, full-field coverage for both resolution modes. 

 \section{Science Case}\label{sec2}

 As stated in the MSE Science Case 2019,  \citep{https://doi.org/10.48550/arxiv.1904.04907}, MSE will enable spectroscopic surveys with unprecedented scale and sensitivity by collecting millions of spectra per year down to apparent magnitudes of m$\sim$24 mag (in low/moderate resolution mode) and m$\sim$20 mag (in high resolution mode).  MSE will elucidate the composition and dynamics of the faint Universe and impact nearly every ﬁeld of astrophysics across a range of spatial and mass scales, from individual stars to the largest scale structures. Key science programs that MSE will execute include the ultimate follow-up of Gaia to understand the chemistry and dynamics of the Milky Way, plus the full extent of the disk and Galactic halo at high spectral resolution; the revolutionary study of galaxy formation and evolution at cosmic noon, conducted at the peak of the star formation history of the Universe; the determination of the neutrino mass; and, the generation of insights into inﬂationary physics via cosmological redshift surveys that probe a large volume of the Universe with a significantly high galaxy density. The data generated by MSE will maximize the science return of an array of ground-based and space-based facilities.  Specifically, MSE will both complement and extend the data from the Gaia Mission \citep{2016A&A...595A...1G}, the Vera C. Rubin Observatory \citep{2019ApJ...873..111I}, the Square Kilometer Array \citep{2009IEEEP..97.1482D}, the Euclid Mission \citep{2011arXiv1110.3193L}, the Nancy Grace Roman Space Telescope \citep{2015arXiv150303757S}.  MSE can additionally serve as a feeder to the 30-meter class telescopes.

As a fully dedicated spectroscopic survey facility with a 12.5m aperture and a 1.5 square degree ﬁeld of view, MSE is designed for transformative, high-precision studies of faint astrophysical phenomena. Approximately 12,000 ﬁbers will feed spectrographs operating at low/moderate spectral resolution (R$\sim$ 3000, 6000) and roughly 6,000 ﬁbers will feed spectrographs operating at high resolution (R$\sim$ 30000).  Focal plane sharing will occur with the low/moderate and high resolution modes being available simultaneously (the entire 360–1800 nm wavelength region will be available at lower resolutions while windows in the optical will be accessible at the highest resolution).
The MSE system is optimized for high throughput observations of very faint sources in the Universe.  The dedicated operational mode of the facility as well as robust calibration and data quality steps ensure that the equivalent of more than 40 million ﬁber hours of 10-m class spectroscopy are available for science every year. Accordingly, MSE will be able to produce datasets equivalent in number of objects to an SDSS Legacy Survey every several weeks, notably on a telescope with a diameter five times larger.
\begin{figure*}[th!]
\centerline{\includegraphics[width=12cm]{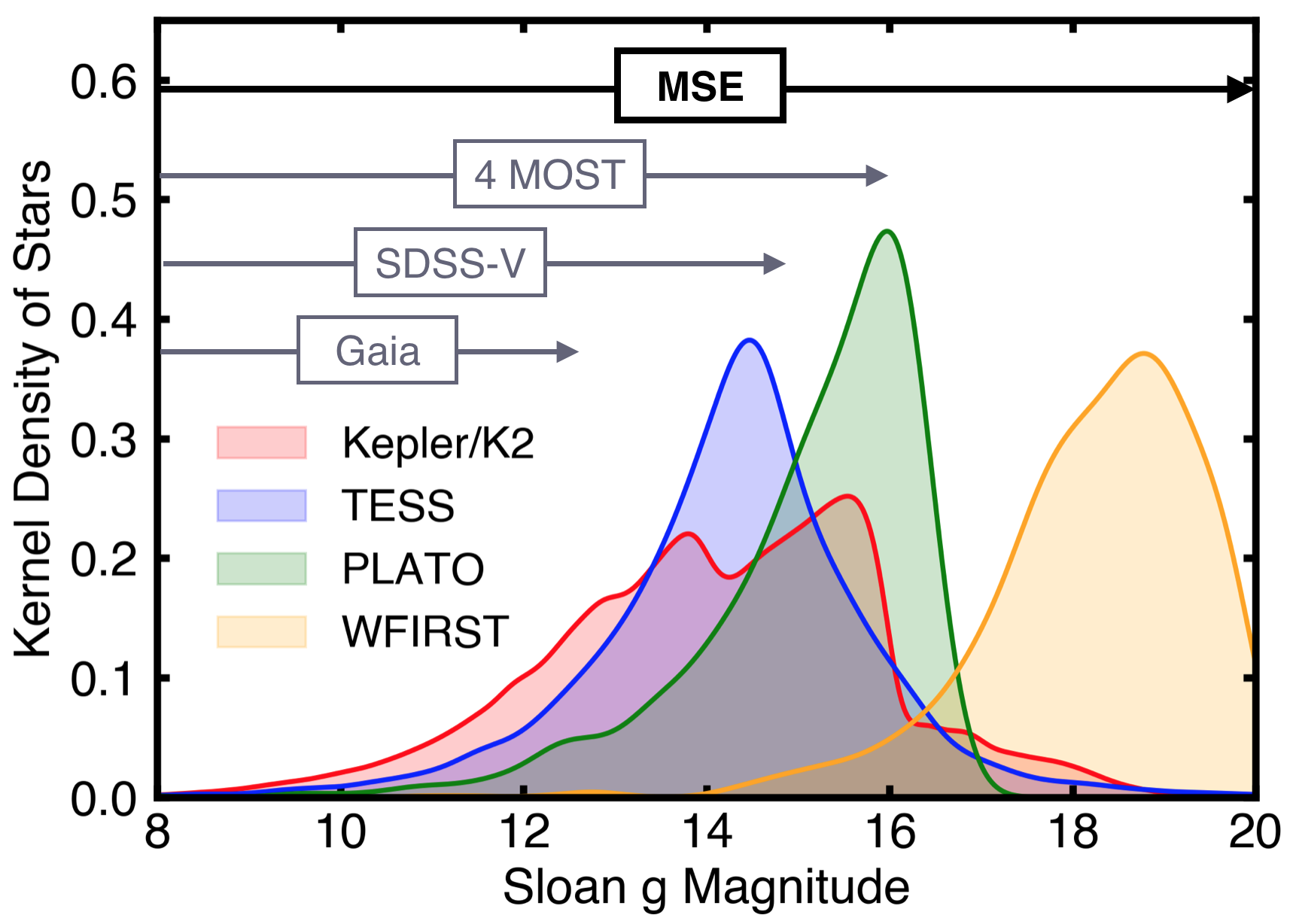}}
\caption{MSE Science Case: Exoplanets and Stellar Astrophysics. MSE will provide high-resolution optical spectroscopy for tens of millions of stars with high-precision, space-based photometry. Lines on this figure indicate the g-magnitude distribution for stars with space-based photometry from Kepler/K2 (red, e.g., \citealp*{2016ApJS..224....2H}) and predicted yields of stars observed with a photometric precision better than 1 mmag hr-1 from an all-sky survey with TESS (blue, e.g., \citealp*{2019AJ....158..138S}), a typical PLATO ﬁeld (green, \citealp*{2014ExA....38..249R}), and the WFIRST/Roman microlensing campaign (orange, e.g., \citealp*{2015JKAS...48...93G}). Shown in gray are the sensitivity limits for other, external MOS facilities that will provide high-resolution (R > 20K) spectroscopy over at least half of the sky. Featured in black is the sensitivity limit for MSE (surpassing the performance of the others). \label{fig2}}
\end{figure*}

Regarding stellar scales and the Milky Way, MSE will be one of the most powerful facilities to perform spectroscopic observations for stars {\it across} the Hertzsprung-Russell diagram. MSE stellar monitoring programs will dramatically improve understanding of stellar multiplicity, providing detailed information for exoplanet host stars as well as dynamical masses for unprecedented samples of high-mass exoplanets (see Figure~\ref{fig2}). MSE will determine the s- and r-process element abundance distributions across the Galaxy as well as in several nearby MW satellites.  It will produce an unrivaled dataset of the most chemically primitive stars with which to identify the signatures of the very ﬁrst supernovae and chemical enrichment events in the Universe. MSE will conduct the quintessential spectroscopic follow-up of the Gaia mission, producing high resolution spectral data products for all Gaia stars northward of -50 degrees declination. Critical to the understanding of the faint and distant regimes of the Milky Way, MSE will carry out an in-situ chemodynamical analysis of millions of individual stars in all Galactic components. It will also play a key role in the unparalleled, three-dimensional ISM mapping efforts. Finally, MSE will usher in a new era for the analyses of Milky Way satellite galaxies.  It will enable robust chemo-dynamical measurements across their full luminosity range as well as provide spectra for over an order of magnitude more stars in each system than is currently achievable.

\begin{figure*}[th!]
\centerline{\includegraphics[width=18cm]{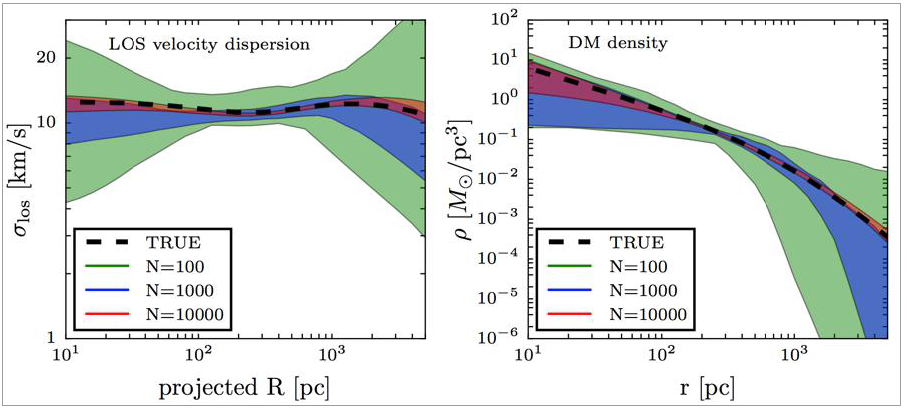}}
\caption{MSE Science Case: Astrophysical Tests of Dark Matter. Recovery of intrinsic line-of-sight velocity dispersion (left) and inferred dark matter density (right) proﬁles as a function of spectroscopic sample size. Shaded regions represent 95 credible intervals from a standard analysis (based on the Jeans equation) of mock data sets consisting of line of sight velocities for N = 102, 103, and 104 stars (median velocity error 2 km s-1), generated from an equilibrium dynamical model for which true proﬁles are known (thick black lines, which correspond to a model having a cuspy NFW halo with $\rho$ (r) $\propto$ r-1 at small radii). \label{fig3}}
\end{figure*}
In the dark sector, MSE will undertake a series of surveys that provide critical input into determinations of the mass function, phase-space distribution, and internal density proﬁles of dark matter halos across all mass scales. Recent N-body and hydrodynamical simulations of cold, warm, fuzzy, and self-interacting dark matter suggest that non-trivial dynamics in the dark sector could have left an imprint on structure formation (see Figure~\ref{fig3}). The unprecedented kinematic datasets from MSE will be used to search for deviations away from the prevailing model in which the dark matter particle is cold and collisionless.

On the scales of galaxies and galaxy groups, MSE will conduct revolutionary extragalactic surveys at the peak of cosmic star formation. At low redshift, MSE will examine a representative volume of the local Universe to stellar and halo masses lower than what is achievable with any current surveys. It will measure the extension of the stellar mass function to masses below 10$^{8}$M$_{sun}$, for a cosmologically representative, unbiased, and spatially-complete spectroscopic sample. While at high redshift, MSE will generate a high-completeness, magnitude-limited sample of galaxy redshifts spanning the epoch of peak cosmic star formation (1.5 < z < 3.0). MSE surveys will have sufficient areal coverage, depth, and temporal character to span the AGN menagerie in this redshift range, and will probe the growth of supermassive black holes (SMBHs) by measuring luminosity functions, clustering, outﬂows, and mergers. A multi-epoch AGN reverberation mapping campaign by MSE will yield 2000-3000 robust time lags, an order of magnitude more than the expected yields from current campaigns.  This MSE data set will enable accurate SMBH mass measurements for one of the largest samples of quasars to date and unprecedented mapping of the central regions.

Regarding the largest scales, MSE will be able to address two of the most important questions within Physics: the neutrino mass and the physics of inﬂation.  It will do so by conducting a cosmological redshift survey that will probe a large volume of the Universe with a high galaxy density. MSE will measure the level of non-Gaussianity as parameterized by the local quantity f$_{NL}$ to a precision of $\sigma$(f$_{NL}$) = 1.8. These MSE data combined with that from a next-generation Cosmic Microwave Background (CMB) experiment will provide the ﬁrst 5-sigma conﬁrmation of the neutrino mass hierarchy (as derived from astronomical observations).

 \section{Instrument Overview}\label{sec3}

\subsection{Quad-Mirror Telescope}

MSE was previously baselined as an 11-meter prime focus telescope with a multi-element and translating corrector/ADC (Atmospheric Dispersion Compensator).  Further detailed examination of the concept revealed optical ghosting issues from the correcting optics that would seriously impact the quality of the faintest spectra targeted by the telescope.  A trade study was explored that included a modified prime focus design, a two-mirror or dual-mirror Cassegrain focus design, and a four-mirror or quad-mirror (QM) anastigmatic design based upon the popular Paul-Baker family of anastigmatic telescopes.  The QM design clearly provided the best mitigation of optical ghosting due to a lower number of transmissive elements, although the modified prime focus and the dual-mirror designs were able to meet acceptable levels of performance.

The QM design  has been selected as the new MSE baseline concept primarily because of the considerable increase in versatility provided by the concept over that offered by either the prime focus or dual-mirror concepts in addition to the superior optical ghost mitigation.  It is anticipated that the QM design will meet all of the MSE objectives.

Rather than having three mirrors as in the Rubin Simonyi Survey Telescope \citep{2019ApJ...873..111I}, the QM design forms an internal focus between M2 and M3.  M3 then images a pupil back at this intermediate focus.  Placing M4 at that pupil allows that mirror to redirect the light to the side of the telescope where spectroscopic instrumentation can more easily be accommodated.  The MSE QM concept pushed this design to have this external focus not far in front of the primary where the elevation axis could be co-located, creating a Nasmyth location for the instrumentation.  This concept is a derivative of some earlier concepts developed as case studies for wide field of view (1 degree) spectroscopy on a 30-meter telescope \citep{2000SPIE.4004..397B} and \citep{2004SPIE.5489..454B}.

\begin{figure*}[th!]
\centerline{\includegraphics[width=15cm]{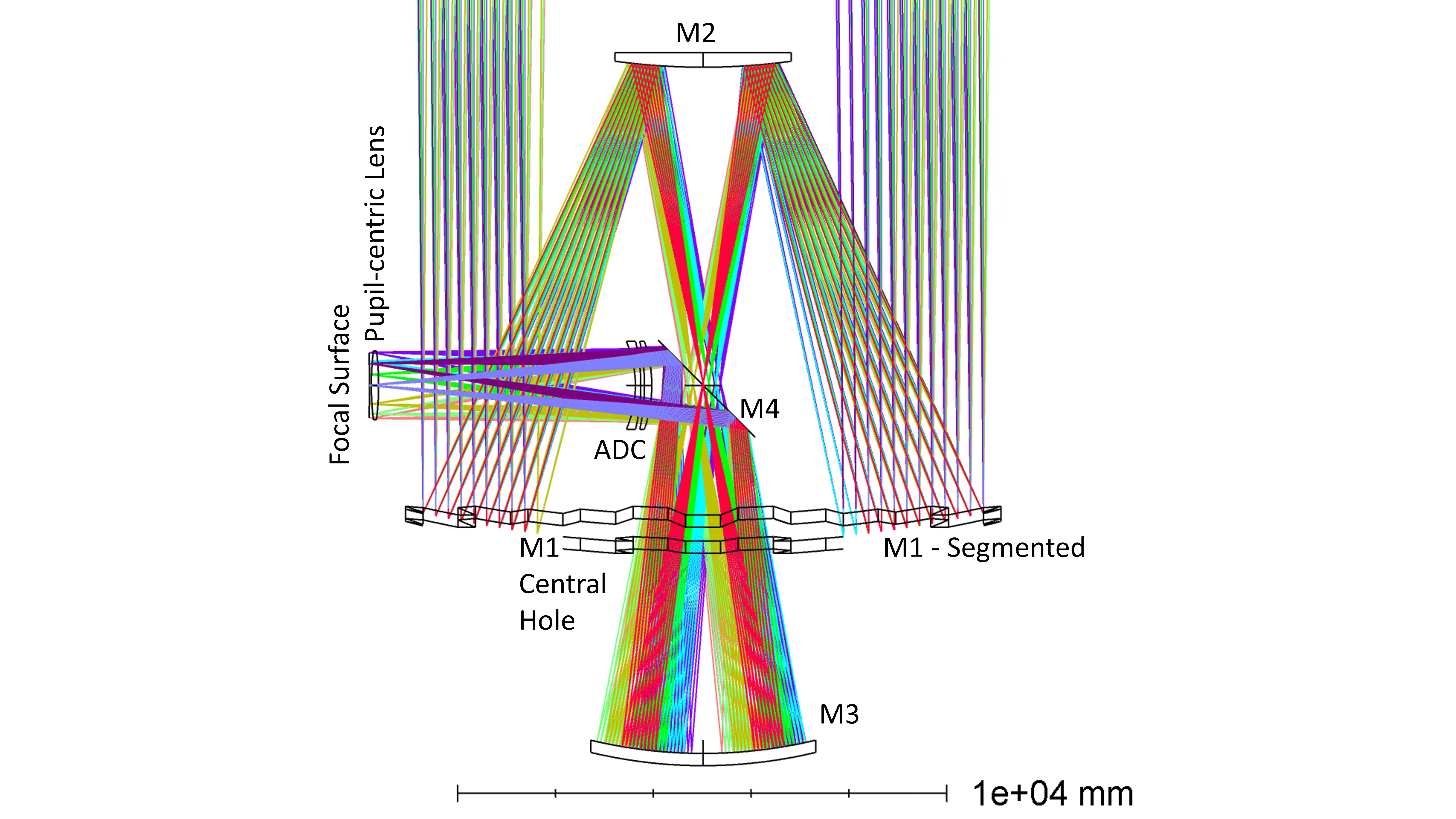}}
\caption{Schematic of the raytrace for the QM Design. \label{fig 4}}
\end{figure*}

A schematic raytrace view of the MSE QM concept is shown in Figure \ref{fig 4}.  The primary mirror (M1) is a nearly paraboloidal (k=-0.88) segmented mirror with a fully circumscribed aperture of 12.5 meters.  M2 is a hyperboloidal mirror (k=-6.3) with a full diameter of 3.6 meters and a large central shadow which could allow lightweighting of the mirror by making it an annulus.  M3 has a full diameter of 4.6 meters and is ellipsoidal (k=-0.37).  Only a small central hole can be permitted in M3 for handling and alignment purposes.  M4 is an elliptical fold mirror 1.9 by 2.8 meters.  M4 must have a central hole to allow the intermediate focus passage from M2 to M3.  The field of view is limited by the size of this central hole.

M2, M3, and M4 are envisioned to be monolithic mirrors as segment production constraints are inadequate.  The diameters of these mirrors must also be no larger than 4.6 meters imposed by the mirror handling/re-coating constraints and the plan to re-purpose the existing CFHT building structure for MSE.

As M3 forms an image of M1 onto the M4 mirror, it may be possible to make M4 adaptive for ground layer image correction over much of the field of view.  Preliminary modeling suggests that worthwhile improvement in the image performance will be possible (work in progress). This will be left as a future possible upgrade for the MSE facility. 

MSE has been granted access to the ELT M1 mirror intellectual property (IP). The expectation is that the MSE M1 will be fabricated using that IP.  The ELT mirror segment technology imposed further constraints on the design.  Peak-to-valley (PV) asphericity for a given segment is limited by the interferometric testing process.  This imposed constraints on both the radius of curvature and the conic/aspheric values for the M1 primary mirror.  Additionally, current segment mounting designs place an additional constraint on the minimum radius of curvature.

The MSE QM design easily fits within the volume designed for the baseline prime focus concept.  Initial mass estimates suggest that it is likely to be less massive than the very long prime focus alternative which requires significant deadweight to counterbalance the bottom-heavy telescope.

The advantages of having a Nasmyth focus are:
\begin{enumerate}
\item Only radial gravitational variations on the focal surface and fiber positioner as the instrument rotates to track the field.
 \item The ability to rotate M4 to illuminate the other Nasmyth port allowing contemporaneous imaging, IFU observations, or other data collection at that port.
 \item The ability to rotate M4 to illuminate smaller focal stations mounted on the telescope.
 \item Possibility of replacing the fiber positioner with some future multi-object spectroscopic instrumentation technology, perhaps micro-shutter fed spectrographs.
\item Shorter fiber feeds compared to the prime focus and Cassegrain configurations.   
\end{enumerate}

The ADC elements are fused silica and very large, nearly 1.8 to 1.9 meters in diameter.  The surfaces are all spherical and both lenses have a bit of wedge.  The lenses are counter-rotated to compensate for the atmospheric dispersion. Fortunately, the location of the lenses results in only radial gravitational variations in their loading.  Also, the large central obstruction of the telescope could allow central axial supports on both lenses.  The need for only fused silica offers the possibility of extending the spectral range of the telescope further into the blue (due in part to the shorter fiber run) and the use of UV-optimized silica for the ADC and field lens.  Data acquisition into longer wavelengths (K-band and beyond) could be possible by folding the beam to a second Naysmyth-like focus, without an ADC and utilizing IR-optimized silica for the field flattener.

In order to extend the focus outside of the incoming light to the telescope, the focal ratio needs to be somewhat slower than f/2.  The current design is f/4.  Although there will be a bit worse focal ratio related etendue loss, the factor of 4 increase in multiplex capability resulting from the larger plate scale will far outweigh the few percent loss of light.  The rather large central shadow in the QM design that is typical of 3+ mirror anastigmats will also mitigate light loss if the spectrographs contain large central obstructions due to, for example, large Schmidt camera systems.

\begin{figure*}[th!]
\centerline{\includegraphics[width=15cm]{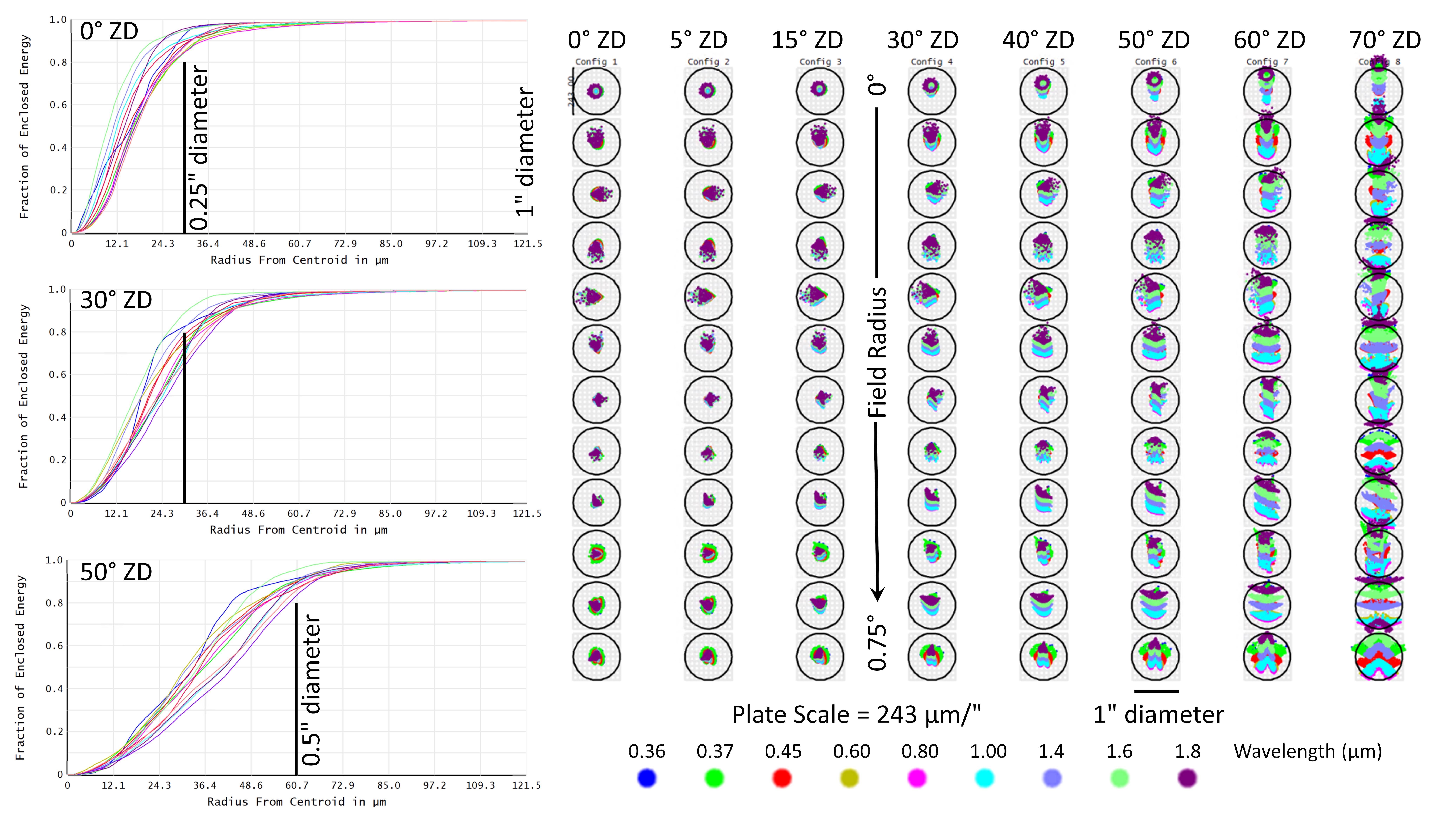}}
\caption{Encircled Energy (EE) and spot diagrams for the QM design. \label{fig 5}}
\end{figure*}

The image quality of the QM telescope is quite good. Figure \ref{fig 5} shows polychromatic (0.36 to 1.8 microns) encircled energy plots indicating that the ideal design delivers 0.25” diameter >80 percent encircled energy at Zenith distances up to about 30 degrees.  >80 percent encircled energy is achieved at a Zenith distance of 50 degrees.  A matrix of spot diagrams is also given in the figure as a function of Zenith distance and radial position within the field for the full range of wavelengths.  These diagrams indicate the quite good performance of the ADC.

\subsection{Fiber Positioners and Fiber Cable}
The current baseline concept for a fiber positioner is the tilting spine technology pioneered at the Australian Astronomical Observatory \citep{2018SPIE10702E..1MS} and soon to be implemented for the 4MOST facility \citep{2014SPIE.9151E..1XS}.  It is not clear how such a fiber positioner will scale with the factor of 4 increase in the physical area of the focal surface.  We estimate that the fiber count will be around 18000 and possibly as high as 20000.  Fibers of 1 arc-second diameter are expected to be used for the low and moderate resolution spectrographs while 0.8 arc-second diameter fibers are anticipated for the high resolution spectrographs.

\begin{figure*}[th!]
\centerline{\includegraphics[width=9cm]{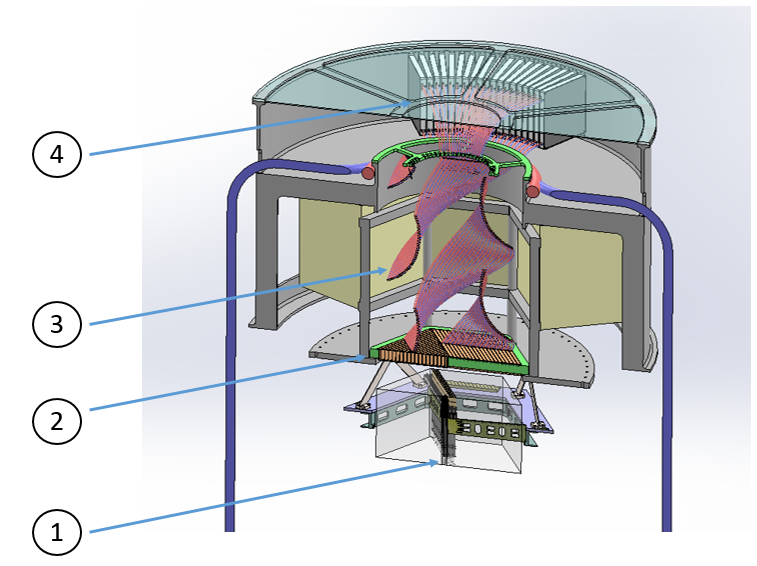}}
\caption{Diagram of the FTS with the following, labeled components: 1) PosS (simplified), 2) Fiber Combiner, 3) Helical Tubes, and 4) Loop Boxes. \label{fig 6}}
\end{figure*}

 All fibers are simultaneously positioned in parallel to maximize the amount of light entering each of them based on individual sky targets. This positioning is performed by actuators in the Fiber Positioner System (PosS). The Fiber Transmission System (FTS) then transmits the light from the focal surface at the top end of the telescope, through the observatory to banks of spectrographs several tens of meters away. Two sets of spectrographs are planned, the Low-Moderate Resolution (LMR) and High Resolution (HR) spectrographs. 
The baseline Fiber Transmission System (FTS) concept (Figure \ref{fig 6}) provided by Herzberg Astronomy and Astrophysics (HAA) was previously described in \citep{2018SPIE10702E..1LH}, \citep{2018SPIE10702E..7SV}. 

Along with providing the fiber bundles, the FTS subsystem also includes a fiber management system, which routes and protects the fibers through all motions of the telescope in all environmental conditions, as shown in Figure \ref{fig 6}. 

Fiber bundles terminate at slit input units that provide the interface from the fibers to the spectrograph. For the baseline spectrograph designs, the interface to the spectrographs is required to have a spherical shape or “smile” due to the off-axis collimators in both spectrograph designs. The shape of the slit compensates for the optical distortion such that spectra are “flat” (or straight line) when delivered to the spectrograph detectors. It is expected that these will look something like the slit input unit from Hermes \citep{2015JATIS...1c5002S} but with the design features, such as V-grooves and strain-relief proposed by HAA.

The design of FTS is primarily driven by fiber throughput requirements and the need to have stable and repeatable calibration over 24 hours over the full range of pointing motion of the telescope. This includes throughput due to transmission losses (mostly based on fiber length and use of continuous fibers), Fresnel (input and output) losses, and focal ratio degradation (FRD). To provide the highest possible throughput, fibers are provided in a continuous link, using no connectors, all of the way from the focal surface to the spectrograph inputs.   

\subsection{Spectrographs}
Detailed baseline designs for the spectrographs have been developed. These designs were based on the optical requirements from the original prime focus design for MSE. The new QM design provides additional challenges as it represents an increase in etendue, due to the increased aperture size and a significant increase in fiber count and a corresponding increase in the effective number of spectrographs.  The designs for spectrographs are thus in flux. Modifications of the current baseline designs are being evaluated in order to understand if they can be modified to accept the larger aperture from the new QM spectrograph design.

In parallel, the MSE team is evaluating wavelength splitting in order to allow for the development of single-arm spectrographs covering different wavelength bands. This is similar in concept to those being developed for the VIRUS spectrographs \citep{2004SPIE.5492..251H} developed for the HetDex \citep{2011ApJS..192....5A} system on the HET telescope \citep{2011ApJS..192....5A}. This module would take the output of the fibers coming from the telescope and separate the light into the desired wavelength bands and reimage onto fiber feeding single channel spectrographs optimized for each band.  This module would remove the need for the large and expensive dichroics currently within the designed spectrographs.  Such removal should result in better packaging of the spectrographs themselves which will be critically important with four to five times more spectrographs than the current prime focus baseline.  The ability to split out the wavelengths also opens up the prospect of locating the blue channel spectrographs closer to the telescope with considerably shorter fibers.  It is anticipated that there could be a gain of two in transmission at 360 nm with such an arrangement.

In addition to wavelength splitting the team is also evaluating the possibility of pupil slicing in order to significantly relax the constraints on the spectrographs, in particular the high resolution spectrographs.

The substantial increase in the number of fibers in the focal plane to over 18,000 will require a significant increase in the number of spectrographs and the number of detectors. Order-of-magnitude estimates correspond to over 100  (6K x 6K) focal plane arrays without pupil slicing and of order 700 focal plane arrays with pupil slicing. 

Baseline detectors are a combination of CCDs and mercury cadmium telluride (HgCdTe)  detectors. In parallel, the MSE team is working with the Teledyne engineering staff to evaluate the possibility of using CMOS detectors. One significant challenge will be cooling between 100 and 700 individual focal plane arrays.

The following section presents the baseline spectrographs. 
\subsubsection{Low/Medium Resolution Spectrographs}
The Low/Medium Resolution spectrograph design for the visible and NIR spectrographs has been provided by Labortoire d’Astrophysique de Marseilles (LAM), Centre de Recherche Astrophysique de Lyon CRAL. The concept was presented in 2021 and includes splitting the overall concept from 2018 into two different flavors of spectrograph: VIS and NIR, with the  wavelength splitting system allowing for all LMR fibers to feed all wavelengths  This was reported previously at SPIE in 2020 \citep{2020SPIE11447E..95J}.

\begin{figure*}[th!]
\centerline{\includegraphics[width=14cm]{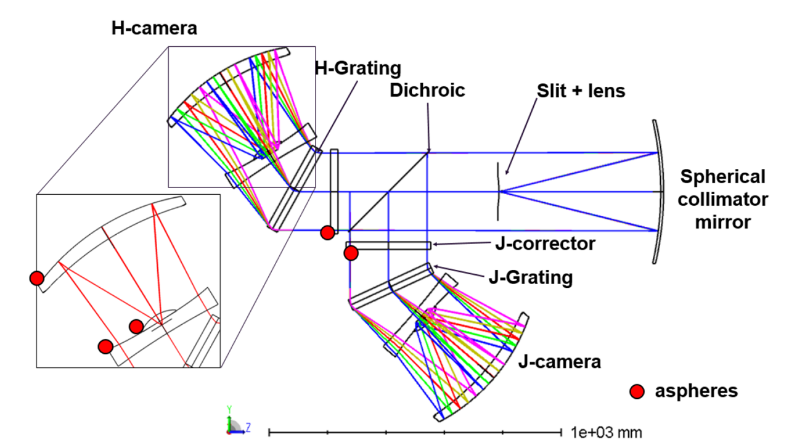}}
\centerline{\includegraphics[width=14cm]{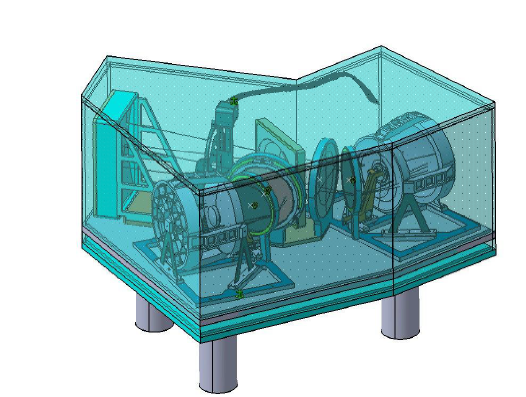}}
\caption{LMR NIR optical layout (top). Optomechanical layout (bottom)\label{fig 7}}
\end{figure*}

The LMR design team developed a NIR design (Figure \ref{fig 7}), inspired by the ELT MOONS “WonderCamera” \citep{2011Msngr.145...11C}. The NIR spectrographs accept ~500 LMR fibers, via the curved slit discussed previously. The catadioptric layout includes two arms,  H-Band and J-Band, an f/0.95 camera, beam sizes approx. 265 mm, and a Mangin mirror. The detectors are two 4k x 4K – 15 um Detectors (H4RG) for each spectrograph. The two NIR units will be in a cryo-cooled environment at 200-220K to reduce thermal background. 

In addition, an optical design for the VIS spectrographs has been proposed by Winlight, in collaboration with LAM, France. The design includes VIS spectrographs that accept > 500 dedicated fibers each, also with a curved slit input. Winlight also adopted a design similar to MOONS. A 4-arm vs. 5-arm architecture with different cameras were compared and it was found that 5-arms is not physically realistic. The chosen design is a 4-arm with VPHG gratings and  max beam diameter of 270 mm. The camera is f/0.9 camera with the filed lens acting as a cryostat window. This design also incorporates 4k x 4K – 15 um Detectors (nominally Teledyne H4RG).

Both designs significantly reduce risk in optical fabrication and are much simpler in that there are fewer components and mechanisms, fewer aspheres, and fewer cryo-cooled components. The design has lower throughput compared to CoDR 2018, approx. 40 percent vs. 50 percent and the impact of this on sensitivity is still being assessed. The designs meet the majority of other requirements except cross-talk between the fibers which is currently being evaluated.
 
These designs are at the concept stage and will be developed further, along with other studies ongoing at MSE.

\subsubsection{High Resolution Spectrographs}
\begin{figure*}[th!]
\centerline{\includegraphics[width=14cm]{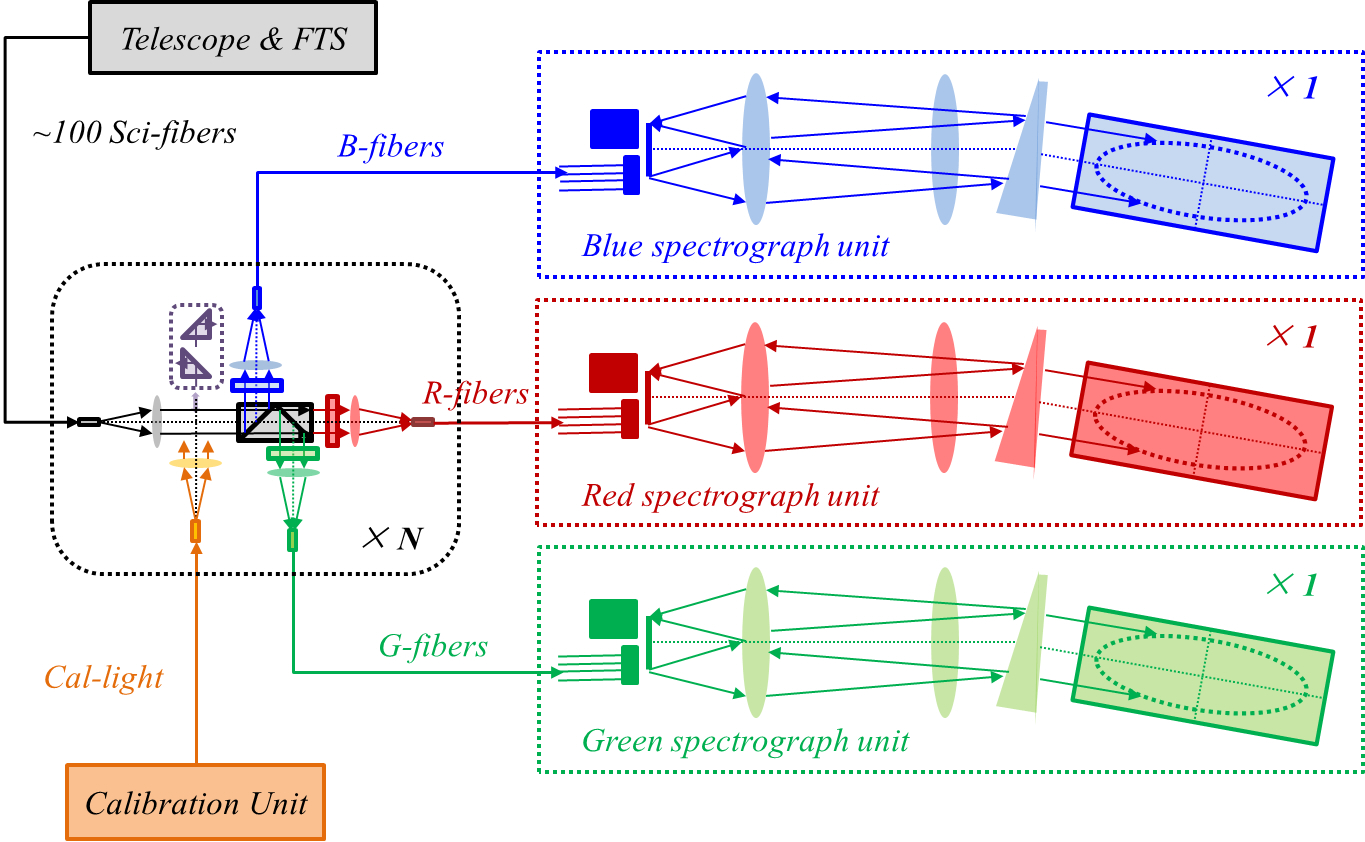}}
\caption{Proposed architecture of the HR Spectrograph with wavelength splitting.\label{fig 8}}
\end{figure*}

After some design iterations and explorations, in 2021, an optical design by NIAOT (Nanjing Institute of Astronomical Optics \& Technology) has been proposed. The details for this design are presented in  \citep{2022SPIE12184E..7OZ}. The design features a bank of spectrographs, each with ~100 fibers per spectrograph. Each spectrograph has 3 channels (B, G, R) with the wavelength splitting in a “pre-optics” unit as shown in Figure (\ref{fig 8}). 

\begin{figure*}[th!]
\centerline{\includegraphics[width=14cm]{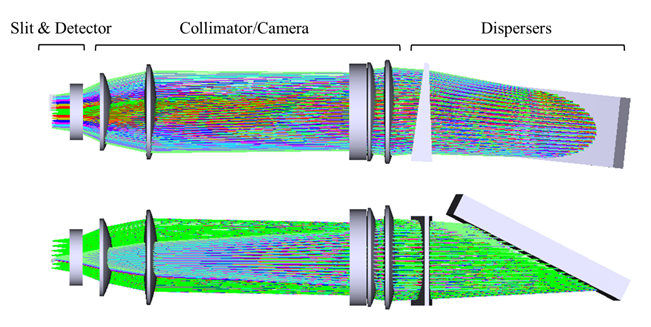}}
\caption{The optical design of the High Resolution Spectrograph.\label{fig 9}}
\end{figure*}

Each preoptics unit includes an 80 um input fiber from FTS and a 120 um output fiber to each channel. Between the input and output fibers, light is split into the blue, green, and red channels using dichroics. Window bandpasses (wavelength windows) are controlled using filters. These filters are selectable on a filter-changing mechanism. As well, the preoptics unit incorporates the injection of a calibration light for wavelength calibration.
Each spectrograph channel (Figure \ref{fig 9}) incorporates a double-pass design with an f/3.12 camera (diameter 285 mm), Echelle grating and three detector sizes, depending on the arm. For the red arm, the detector will likely have to be a mosaic of two 4k x 4k detectors.

The HR concept reduces some of the previous risks in fabrication and procurement, especially by using conventional Echelle gratings instead of high-density gratings. It will be necessary to investigate the cost implications of the many optical elements in the system and the large number of spectrographs as well as the feasibility within the space available in the MSE observatory. As well, there are some areas to explore in the optical design, including ghosts in the double-pass arrangement, stray light due to the filters, and crosstalk between adjacent fibers on the detector.
The wavelength splitting concept will be evaluated in parallel with the concept design development.

 \section{MSE Pathfinder on CFHT}\label{sec6}

 \begin{figure*}[th!]
 \centerline{\includegraphics[width=13cm]{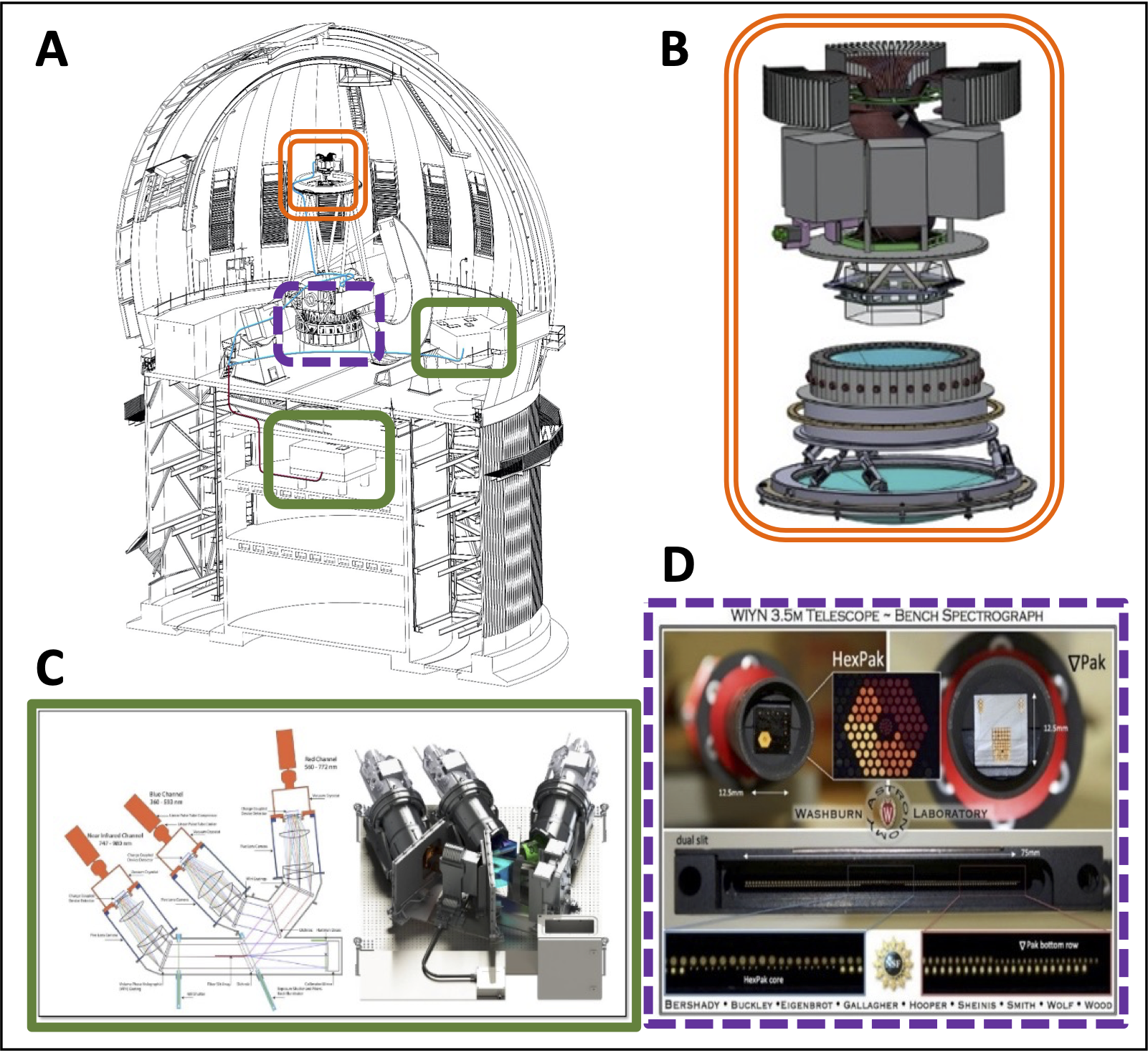}}
\caption{MSE-Pathfinder and representative components. A) CFHT telescope with Pathfinder system. B) Positioner and wide-field corrector (\textit{compound orange border}). C) Bench-mounted spectrographs, e.g., DESI (\cite{2016arXiv161100036D}; \textit{solid green border}). D) WIYN telescope IFU similar to Cassegrain mounted pathfinder IFU (\cite{2012SPIE.8446E..2WW}; \textit{dotted purple border}) \label{fig10}}
\end{figure*}

 MSE/CFHT plan to develop an end-to-end Pathfinder (Figure \ref{fig10}) for the Maunakea Spectroscopic Explorer. Pathfinder will fast-track technology development for MSE by demonstrating on-sky the ability of some of MSE's major components and software packages. In parallel, Pathfinder will produce an initial science product that can be shared with the community, including community access to this wide field spectroscopic capability. Motivation for the Pathfinder also comes from the recent report from the National Academies, “Pathways to Discovery in Astronomy and Astrophysics for the 2020s” (Astro2020). Astro2020 Section 7, “Realizing the Opportunities: Medium and Large-scale programs” calls for three funding tracks to achieve the science priorities the Decadal Review Panel identified. Consistent with these priorities, the primary science goals of the Pathfinder are time-domain astrophysics: specifically spectroscopic follow-up in order of transients identified by facilities such as Rubin Observatory and Zwicky Transient Factory to optimize their identification and classification; Galactic archaeology; and the spectroscopy of stars for stellar abundance studies and stellar evolution studies. The third track calls specifically for upgrades to existing facilities and community access to additional facilities. The Pathfinder instrument is in direct response to the call from track three. 
The end-to-end Pathfinder will include multi-object spectrographs fed by a 1000-fiber positioner covering 1 square degree at prime focus of the CFHT and/or a 1000-fiber IFU covering approximately 33" x 33" FOV from CFHT. It will achieve a spectral resolution greater than R=2000 in the visible with potential upgrades including the NIR (J and H band) and high resolution in the visible. Pathfinder is being developed with low risk and aggressive schedule as a priority. Hence it will not utilize the MSE spectrograph design and instead use existing/proven designs, possibly similar to the DESI spectrographs. \citep{2015AAS...22533608E}. The Pathfinder project will develop the prototype software architecture for MSE, including scheduling, targeting, data reduction and analysis, and data management and archival. 

 



\section{Conclusions}\label{sec5}

MSE will be a new massively multiplexed, wide field spectroscopic survey facility located on the site of the Canada-France-Hawai\okina i Telescope and reusing the existing footprint and much of the existing building structure and infrastructure. This paper has presented an overview of scientific motivation as well as a description of the major technical subsystems of the facility. The full facility will come online in the late 2030s and a Pathfinder instrument, currently being developed for CFHT will come online in the late 2020s.


\section*{Acknowledgments}
The Maunakea Spectroscopic Explorer project acknowledges the reverence and importance that the summit of Maunakea holds within the Native Hawaiian community. Maunakea is a place of worship, the realm of deities, where earth and sky connect. There are hundreds of historic properties, archaeological remains, shrines, and burial sites on its slopes and summit. As astronomers, we are privileged and honored to have the opportunity to observe the sky and advance our knowledge of the universe through the facilities hosted there.
 
The Canada-France-Hawai\okina i Telescope, its funding organizations, and its contributing partners support the values of community-based astronomy promoted in the Astro2020 report. MSE is grounded in the belief the future of astronomy on Maunakea requires a new model, one which fully involves the local and Native Hawaiian communities in the governance of the mountain and in the definition of new projects at their inception. Together with the Hawai\okina i Island community and Mauna Kea Stewardship and Oversight Authority, we will define key measurable goals for future CFHT development in terms of local employment, governance, and sustainability as an integral part of the MSE-Design Study and MSE-Pathfinder.



%

\bibliography{Wiley-ASNA}

\end{document}